\documentstyle[11pt,aaspp4]{article}

\begin{document}

\title{Measuring the Virial Temperature of Galactic Halos Through Electron
Scattering of Quasar Emission Lines} 
\author{Abraham Loeb}
\medskip
\affil{Astronomy Department, Harvard University, 60 Garden Street, Cambridge,
MA 02138}


\begin{abstract}

Semi-analytic models of galaxy formation postulate the existence of
virialized gaseous halos around galaxies at high redshifts.  A small
fraction of the light emitted by any high-redshift quasar is therefore
expected to scatter off the free electrons in the halo of its host galaxy.
The broadening of the scattered emission lines of the quasar can be used to
measure the temperature of these electrons. For gas in virial equilibrium,
the velocity width of the scattered line profile is larger by a factor of
$\sim (m_p/m_e)^{1/2}=43$ than the velocity dispersion of the host galaxy
and reaches $\ga 10,000~{\rm km~s^{-1}}$ for the massive galaxies and
groups in which bright quasars reside. In these systems, the scattered
width exceeds the intrinsic width of the quasar lines and hence can be used
to measure the virial temperature of the quasar host.  The high degree of
polarization of the scattered radiation can help filter out the extended
scattered light from the central emission by the quasar and its host
galaxy. The signal-to-noise ratio of the spectral broadening can be
enhanced by matching the full spectrum of the scattered radiation to a
template of the unscattered quasar spectrum.  Although the central fuzz
around low-redshift quasars is dominated by starlight, the fuzz around
high-redshift quasars might be dominated by scattering before galaxies have
converted most of their gas reservoirs into stars.

\end{abstract} 


\section{Introduction}

The depth of galactic potential wells has been traditionally inferred by
dynamical methods (HI or stellar rotation curves and satellite dynamics) or
gravitational lensing. Unfortunately, the application of these methods to
galaxies at very high redshifts proves difficult.  The resulting
uncertainty in the velocity dispersion of galactic halos prevents current
data sets on the abundance and clustering of galaxies at $z\sim 3$ from
being able to discriminate among different cosmological models (Adelberger
et al. 1998).

Semi-analytic models of galaxy formation and evolution (e.g., Kauffmann,
White, \& Guiderdoni 1993; Baugh et al. 1998; Somerville, Primack, \& Faber
1998) postulate the existence of gaseous halos at the virial temperature
around galaxies at high redshifts. The halos are expected to contain gas
that was assembled into the galaxies but was unable to cool between
successive heating episodes, caused by either merger events or 
supernova-driven winds. The supply of cold gas from the dense core of these
halos controls the star formation rate of their host galaxies.  Although
X-ray halos are rare among local galaxies (Bregman \& Glassgold 1982;
Fabbiano 1989), they might have been prominent during their early
evolution. A direct measurement of the gas temperature in these halos could
determine the potential depth of their host galaxy.

In this {\it Letter} we propose a novel method for probing the virial
temperature of galactic halos, in those cases where a bright quasar shines
at their center.  The quasar radiation is scattered by the free electrons
in the halo and the scattered emission lines are Doppler broadened by the
thermal velocity of these electrons\footnote{Interestingly, this broadening
effect was only recently detected in a local astrophysical setting -- the
solar wind, by the SOHO satellite (Fineschi et al. 1998).}. Due to the
small electron mass, the velocity width of the scattered lines is larger by
a factor of $\sim (m_p/m_e)^{1/2}=43$ than the velocity dispersion of the
galaxy, and could reach values $\ga 10,000~{\rm km~s^{-1}}$ for the massive
galaxies and groups in which bright quasars reside. In these objects, the
scattered line width exceeds the typical intrinsic width of the quasar
emission lines (FWHM$\sim 5,000~{\rm km~{s^{-1}}}$; see, e.g. Peterson 1997
for a review) and can be used to measure the electron temperature.  Since
quasars tend to reside in dense environments such as groups of galaxies
(Hartwick \& Schade 1990; Bahcall \& Chokshi 1991; Ellingson, Green \& Yee
1991; Fisher et al. 1996), additional scattering may also result from the
intra-group gas.

In \S 2 we calculate the expected shape and amplitude of the scattered
quasar lines as a function of the gas temperature and density
profile. Finally, \S 3 discusses the implications of our results.

\section{Scattered Line Profile}

The redistribution function $R(\nu^\prime,{\bf n}^\prime; \nu,{\bf n})
d\nu^\prime d\nu (d\Omega^\prime/4\pi) (d\Omega/4\pi)$ provides the
probability of scattering a photon from a frequency between
$(\nu^\prime,\nu^\prime+d\nu^\prime)$ and solid angle $d\Omega^\prime$ in
the direction ${\bf n}^\prime$ to a frequency between $(\nu,\nu+d\nu)$ and
solid angle $d\Omega$ in the direction ${\bf n}$. For Thomson scattering by
a Maxwellian population of electrons at a temperature $T$, the
redistribution function is given by (Dirac 1925; Mihalas 1978),
\begin{equation}
R(\nu^\prime,{\bf n}^\prime; \nu,{\bf n})={3\over 4}(1+\mu^2){1\over 
\left[2\pi \beta_{T}^2(1-\mu)\nu^{2}\right]^{1/2}} \exp\left[-{
(\nu-\nu^\prime)^2\over 2\beta_T^2 (1-\mu)\nu^{2}}\right],
\label{eq:R_e}
\end{equation}
where $\mu={\bf n}^\prime\cdot {\bf n}= \cos \theta$, and
\begin{equation}
\beta_T\equiv \left({2kT\over m_ec^2}\right)^{1/2},
\label{eq:beta}
\end{equation}
is the typical thermal velocity of an electron in units of the speed of
light $c$. Here $m_e$ is the electron mass, and $k$ is Boltzmann's
constant. The redistribution function can be integrated over any arbitrary
scattering geometry to obtain the profile of a scattered quasar line. 

As an illustrative example, we consider below the case of a spherical,
optically--thin, gas cloud which scatters a narrow line of frequency
$\nu_0$ and total luminosity $L_0$ from a central source.  We assume that
the source luminosity per unit frequency is
$L_{\nu^\prime}=L_0\delta(\nu^\prime-\nu_0)$. An electron at a radius $r$
scatters a net fraction $(\sigma_{\rm T}/4\pi r^2)$ of $L_0$, where
$\sigma_{\rm T}=0.67\times10^{-24}~{\rm cm^2}$ is the Thomson
cross-section. Hence, the scattered spectral luminosity at a frequency
$\nu$ is
\begin{equation}
 L_\nu= \int d^3r~n_{\rm e} \left({\sigma_{\rm T}\over 4\pi r^2}\right)
\int {d\Omega \over 4\pi} \int {d\Omega^\prime \over 4\pi} 
\int d\nu^\prime~R(\nu^\prime,{\bf n}^\prime; \nu,{\bf n}) 
L_{\nu^\prime}(\nu^\prime),
\label{eq:L_nu}
\end{equation}
where $n_{\rm e}(r)$ is the number density of electrons at a radius $r$.
By substituting $d^3r=4\pi r^2 dr$ and
$L_{\nu^\prime}=L_0\delta(\nu^\prime-\nu_0)$ and integrating by parts, we
get (Rybicki \& Hummer 1994)
\begin{eqnarray}
{\beta_T \nu_0 L_\nu\over \tau L_0}&=& \beta_T \nu_0 \int_{-1}^{1}
{d\mu\over 2} R(\nu_0,\nu,\mu) \nonumber \\ &=& 
\left[\left({11\over 10}+{2\over 5}\Delta^2+{\Delta^4\over 20}
\right){e^{-\Delta^2/4}\over {\sqrt{\pi}}} -\left({3\over 2}+{\Delta^2\over
2}+{\Delta^4\over 20}\right){\vert \Delta\vert \over 2} {\rm
erfc}\left({\vert \Delta\vert \over 2}\right)\right] ,
\label{eq:profile}
\end{eqnarray}
where, ${\rm erfc}(x)\equiv(2/{\sqrt{\pi}})\int_x^\infty\exp(-t^2) dt$~~is
the complementary error function;
\begin{equation}
\Delta\equiv {(\nu-\nu_0)\over \beta_T \nu_0},
\label{eq:delt}
\end{equation}
is the frequency shift relative to the line center in units of the typical
Doppler shift of a thermal electron, and
\begin{equation}
\tau=\int_{0}^{\infty} dr~n_{\rm e} \sigma_{\rm T} ,
\label{eq:optical_depth}
\end{equation}
is the radial optical depth of the halo. The integral of the
right-hand-side of equation~(\ref{eq:profile}) over $\Delta$ is unity.  The
scattered profile of a line with a finite initial width can be obtained by
convolving the initial line profile with the kernel given by
equation~(\ref{eq:profile}). This kernel is plotted as the solid line in
Figure 1.

The effective width of the scattered line can be found from the
second-moment, $\langle\Delta^2\rangle= (\int \Delta^2 L_\nu d\Delta)/\int
L_\nu d\Delta$. We find,
\begin{equation}
{\delta \nu\over \nu}\equiv \langle \left({\nu -\nu_0\over
\nu_0}\right)^2\rangle^{1/2}=  \beta_T,
\label{eq:delta}
\end{equation}
where the angular brackets denote an average over $L_\nu$.

Next, we apply the above results to the particular halo profile of an
isothermal sphere with a one-dimensional velocity dispersion
$\sigma_v$. For a fully-ionized gas of electrons and protons, $kT=m_p
\sigma_v^2$, where $m_p$ is the proton mass.  We then get from
equation~(\ref{eq:delta}),
\begin{equation}
{\delta \nu\over \nu}c=1.2\times 10^4 \sigma_{200}~~{\rm km~s^{-1}} ,
\label{eq:width}
\end{equation}
where $\sigma_{200}=(\sigma_v/200~{\rm km~s^{-1}})$.  We assume that the
gas provides a fraction $f_{\rm gas}$ of the total mass and follows the
radial profile of the dark matter at large radii, but has a core of radius
$r_{\rm c}$. Hence, we take $n_{\rm e}=f_{\rm gas}\sigma^2/[2\pi G m_p
(r_{\rm c}^2+r^2)]$, where $G$ is Newton's constant.
Equation~(\ref{eq:optical_depth}) then yields the net fraction of the
quasar light that gets scattered,
\begin{equation}
\tau=2\times 10^{-2} f_{10} \sigma_{200}^2 r_1^{-1},
\label{eq:tau}
\end{equation}
where $f_{10}\equiv (f_{\rm gas}/10\%)$ and $r_1\equiv (r_{\rm c}/1~{\rm
kpc})$. The value of $f_{\rm gas}$ can be as high as $20\%$ if the
cosmological density parameters of baryons and dark matter are 0.06 and
0.3, respectively.  The radius inside which the cooling time due to
free-free emission is shorter than a fraction $\eta$ of the Hubble time is
$\sim 25~{\rm kpc}~(\eta f_{10}\sigma_{200})^{1/2} [(1+z)/4]^{-3/4}$.  The
relevant numerical value of $\eta\la 1$ depends on the average fraction of
the Hubble time that passes between major mergers, as these mergers could
reheat the diffuse gas to its virial temperature. In fact, if quasar
activity is triggered by galaxy mergers, then $\eta$ may be as small as the
ratio between the dynamical timescale of the host galaxy (during which the
quasar is fueled) and the Hubble time, namely $\sim 0.1 (r/30~{\rm
kpc})\sigma_{200}^{-1}[(1+z)/4]^{3/2}$.  Inside the cooling radius, hot gas
may still be replenished by supernova-driven winds (in which case the
geometry of the scattering material is likely to be non-spherical).  The
quasar radiation itself is expected to ionize and heat much of the
quiescent interstellar medium as well, and so overall it is natural to
expect $r_1\la 10$. In resolution-limited observations where the quasar has
to be subtracted from the image, the effective value of $\tau$ would be
dictated by the extent of the quasar mask which sets a minimum value to
$r$.  Scattering by dust would also contribute to the diffuse light, but
would not broaden the quasar lines significantly. For the massive galaxies
and groups in which bright quasars tend to reside (Hartwick \& Schade 1990;
Bahcall \& Chokshi 1991; Ellingson et al. 1991; Fisher et al. 1996), the
width of the scattered emission lines will typically be $\ga 10^4~{\rm
km~s^{-1}}$, i.e. in excess of their intrinsic width.

Note that the optical depth of a cluster of galaxies with $\sigma_v\sim
10^3~{\rm km~s^{-1}}$ and a core radius $r_{\rm c}\sim 250$ kpc is the same
as that of a galaxy with $\sigma_v\sim 200~{\rm km~s^{-1}}$ and $r_{\rm
c}\sim 10$ kpc. However, in the cluster case, the surface brightness of the
scattered light is a factor of $25^2=625$ smaller and is much more
difficult to separate from contaminating background light. Moreover, the
lines scattered by the cluster would only appear as a broad hump on top of
the continuum because of their large width, $(\delta \nu/\nu)\sim 0.2$.
Scattering of beamed quasar light by intracluster gas was suggested as the
origin of the observed alignment between the optical continuum and radio
axes of high redshift radio galaxies (Fabian 1989).

The signal-to-noise ratio for the detection of electron broadening can be
improved by using a full template of the quasar spectrum with all of its
lines, such as O VI, Ly$\alpha$, C IV, Mg II, and
H$\alpha,\beta,\gamma$. The scattered lines will be centered on the
template wavelengths and be broadened by the same amount (except for the
blue wing of the Ly$\alpha$ line which gets absorbed by the Ly$\alpha$
forest of the intergalactic medium). The broadening due to electron
scattering can be distinguished from the intrinsic width of the lines since
it should be the same for all lines while the intrinsic width is often
different for different lines (Peterson 1997).

For a density profile $n_{\rm e}\propto r^{-\alpha}$ the degree of
polarization near the limb of the scattering halo is ${\it
P}=(\alpha+1)/(\alpha+3)$ (Cassinelli \& Hummer 1971). In the singular
isothermal sphere case, $\alpha=2$, and so ${\it P}=0.6$.  Since the
intrinsic quasar emission is often weakly polarized, this high degree of
polarization can be used to filter out the scattered component from the
emission component in the spectrum of the quasar and its host (and may also
be used to reduce the surface brightness contrast between the quasar
pointlike emission and the extended scattered light). Since the net
polarization of a spherical halo is zero, the detection of polarization
requires that the halo be resolved.  At any point, the ${\bf E}$-vector
polarization axis is perpendicular to the radius vector from the halo
center.

In spectroscopic observations for which the halo is resolved, the line
profile might deviate from the shape derived above. In particular, the
spectral shape in equation~(\ref{eq:profile}) holds only if the flux is
summed over a symmetric ring or over half of the symmetric projection of
the entire halo on the sky (with the value of $\tau$ depending on
geometry). If a spectrum is taken for any other portion of the halo, then
the line profile would be different. For example, in the limit of high
angular resolution (needed for the detection of polarization) and a steep
electron density profile (for which the polarization approaches $100\%$),
the line shape would approach $R(\nu_0, \nu)$ for a scattering angle of
$\theta=90^\circ$ ($\mu=0$), since most of the scattered flux would be
contributed from the point of closest approach of the line-of-sight
relative to the source. The dashed line in Figure 1 compares the $90^\circ$
scattering profile of equation~(\ref{eq:R_e}) to the spectral shape given
by equation~(\ref{eq:profile}).

\section{Discussion}

We have shown that the scattered light of a central quasar can serve as a
thermometer of hot galactic halos with a virial temperature $T=5\times
10^6\sigma_{200}^2~{\rm K}$. This probe complements other methods for
measuring the depth of galactic potential wells, although it is limited
only to galaxies which harbor a quasar\footnote{While it is possible also
to search for broadening of stellar emission lines in starburst or normal
galaxies, the scattered flux in these cases would typically be much lower
than for bright quasars.}. Local observations of galactic centers
(Magorrian et al. 1998) and popular models of quasar evolution (e.g.,
Haehnelt \& Rees 1993; Haiman \& Loeb 1998) suggest that all galaxies went
through a quasar phase during their history, and so this method could be
applied to a fair sample of all galaxies.

The scattered flux is expected to replicate the quasar spectrum with
broadened line widths of $\sim 10^4 \sigma_{200}~{\rm km~s^{-1}}$, and to
include a fraction of up to a few percent of the quasar light [see
Eqs.~(\ref{eq:width}) and~(\ref{eq:tau})]. Its contrast relative to the
stellar emission by the host galaxy should increase towards the rest-frame
UV.  Current photometric observations of quasar hosts are sensitive to
scattered light (fuzz) within several arcseconds ($\la 10$ kpc) from the
quasar, at a minimum level of $\sim 1\%$ of the quasar emission (e.g.,
McLeod \& Rieke 1995; Bahcall et al. 1997, and references therein). The
method described here requires low-resolution spectroscopy at a comparable
contrast level.  The signal-to-noise ratio of the spectroscopy can be
improved considerably by filtering out only polarized light from a resolved
halo image. Under more favorable contrast ratios, the feasibility of
detecting extended Ly$\alpha$ emission around high redshift quasars has
already been demonstrated in many cases (e.g., Heckman et al. 1991a,b;
Bremer et al. 1992; Hu, McMahon \& Egami 1996).

The scattering in elliptical galaxies might benefit also from the presence
of a surrounding intra-group or intra-cluster gas.  In spirals, the
scattering by hot supernova-heated gas would add to the more diffuse
scattering in the surrounding halo. The effect of hot interstellar gas in
galactic disks would be heavily contaminated by dust scattering and
absorption and by stellar emission. Although the scattering by dust should
also be polarized, it would not increase significantly the width of the
quasar emission lines. For example, the broad Mg II emission observed from
the extended image of the radio galaxy 3C 265 (Dey \& Spinrad 1996) is most
likely associated with dust scattering of nuclear emission, based on its
flux, polarization, and the constancy of the measured line width as a
function of radius.

We note that Compton heating or cooling by the quasar radiation itself has
only a weak effect on the halo gas over a typical quasar lifetime.  The
characteristic timescale for Compton heating or cooling equals $\sim
3\times 10^8 L_{46}^{-1} r_1^{2}\vert 1- 2T_7/\sigma_{200}^2\vert^{-1}~{\rm
years}$, where $L_{46}$ is the quasar luminosity in units of $10^{46}~{\rm
erg~s^{-1}}$, and $T_7$ is the Compton temperature of the quasar radiation
in units of $10^7~{\rm K}$ (its typical value; see Mathews \& Ferland
1987). This timescale exceeds the Hubble time at $z\sim 3$ for $r\ga 2
L_{46}^{1/2}$~kpc.

The surface brightness distribution of scattered light can be used to infer
the density distribution of the halo gas. When combined with a measurement
of the gas temperature, such data could constrain the mass profile of the
dark matter as well. This method, however, might be compromised by quasar
variability (Wise \& Sarazin 1992). The scattered light from a radius $r$
is delayed relative to the central quasar emission by $\sim 3\times 10^4
(r/10~{\rm kpc})$~years, and so variability of the quasar on this timescale
would affect the brightness distribution of the halo.

Detection of scattered light can also probe beaming in the quasar emission
(Fabian 1989; Sarazin \& Wise 1993).  In the case of beamed {\it broad
line} emission, the inferred electron temperature would be uncertain by a
factor of $(1-\cos\theta)^{-1}$ due to the unknown three-dimensional
orientation of the beam relative to the line of sight. Also, the net
polarization of the scattered light from the entire halo would not vanish
in this case as it does in a spherical geometry. Non-sphericity or
clumpiness in the distribution of infalling gas around the quasar host
(Rees 1988) would result in similar effects.

In cases of beamed quasar emission, the contrast between the nuclear and
host luminosities might be substantially reduced when the quasar beam is
not directed at the observer. Indeed, the possibility that dust or electron
scattering of quasar light accounts for the alignment between the radio and
optical axes of {\it radio galaxies} is supported by observations of some
objects (e.g., di Serego Alighieri et al. 1989; Dunlop \& Peacock 1993;
Cimatti et al. 1993, 1994, 1997; Jannuzi et al.  1995; but see also Dey et
al. 1997; Pentericci et al. 1998). However, only a small minority of all
quasars are radio loud and so these objects are by no means representative
of the entire quasar population.

The existence of hot gas in halos of high-redshift galaxies without quasars
can be probed by complementary measurements in the X-ray and radio bands.
The forthcoming AXAF\footnote{see http://asc.harvard.edu} satellite will
achieve arcsecond resolution and might detect free-free emission from the
brightest halos.  High-resolution radio telescopes, such as the VLA or
MERLIN, could be used to search for the associated Sunyaev--Zel'dovich
effect of these halos and could provide an independent estimate of their
temperature when combined with X-ray imaging data.

Finally, we note that the scattering efficiency of electrons or dust yields
a maximum ratio between the luminosity of quasars and their hosts.
Although the colors and morphology of the fuzz around low-redshift quasars
indicate association with starlight (e.g., McLeod \& Rieke 1995; Bahcall et
al. 1997), scattering may dominate the fuzz around high redshift quasars at
a time when galaxies have not converted yet most of their gas into stars.
It would therefore be interesting to search for an extended scattering
component with quasar colors in the photometry of high-redshift hosts
(e.g., Kukula et al. 1998).  The relative significance of starlight versus
scattered quasar light would indicate whether stars preceded quasars
throughout the entire history of galaxies.

\acknowledgements 

This work was supported in part by the NASA grant NAG5-7039.  I thank
Daniel Eisenstein, Ari Laor, and George Rybicki for useful discussions.

\clearpage
\newpage
\begin{figure}[b]
\vspace{2.6cm} \includegraphics{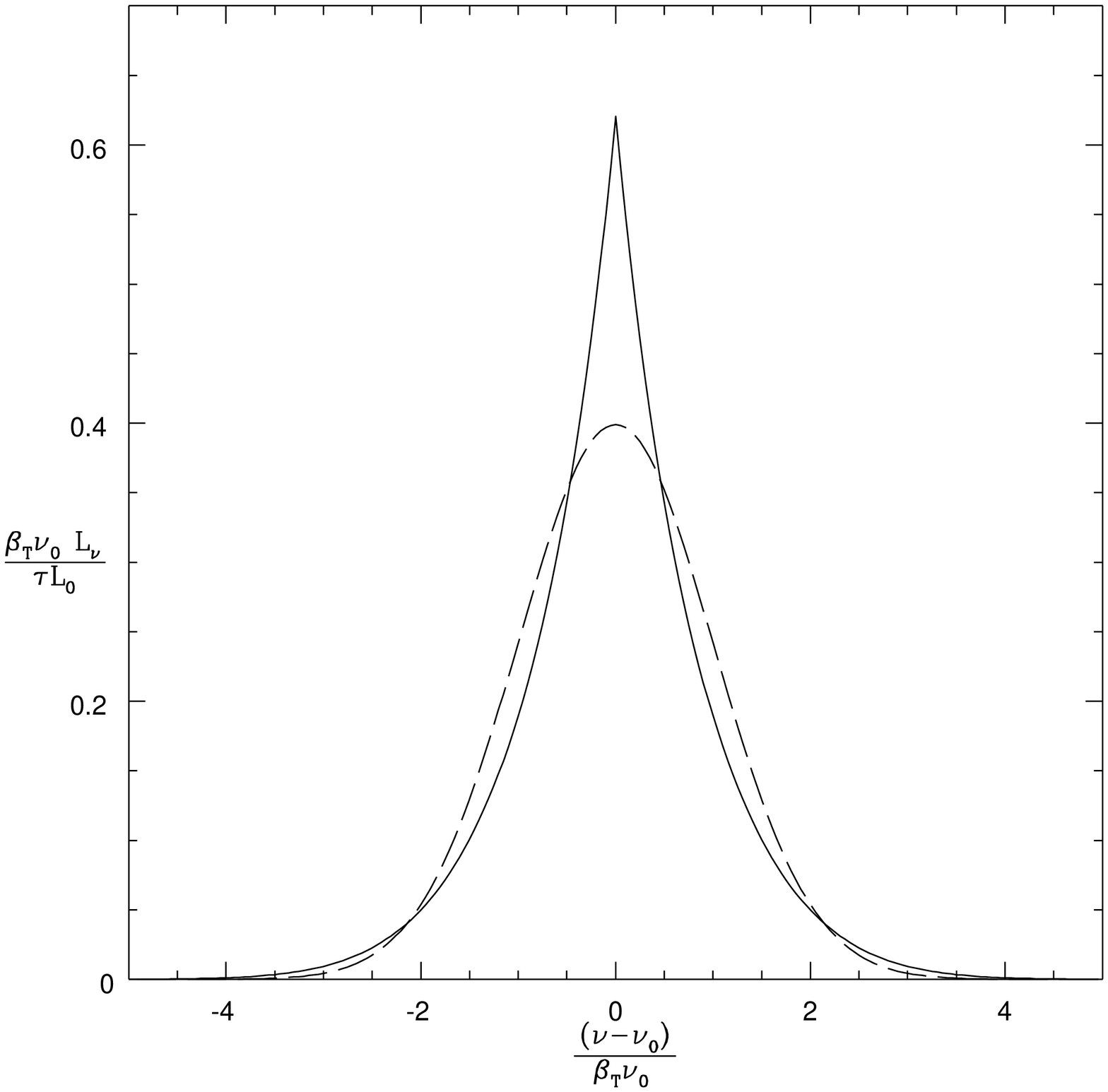}
\vspace*{4.5in}
\caption[Spectrum1] {\label{fig1:sources} Profile of a scattered line from
a spherical halo [solid line; see Eq.~(\ref{eq:profile})] in comparison
with the profile for $90^\circ$ scattering and the same flux normalization
(dashed line). Both profiles have a unit integral and a unit
second-moment.}
\label{fig:1} 
\end{figure}

\end{document}